\documentclass[runningheads]{llncs}
\usepackage[T1]{fontenc}
\usepackage{graphicx}
\usepackage{longtable}
\usepackage{booktabs}
\usepackage{multirow}
\usepackage{todonotes}
\usepackage{subcaption}
\usepackage{comment}
\begin{document}
\title{Large language models for behavioral modeling: A literature survey}

\author{Muhammad Laiq\orcidID{0000-0002-5964-5554}} 
%
%
\institute{Department of Communication, Quality Management and Information Systems \\
Mid Sweden University, Campus Östersund, Sweden \\
\email{muhammad.laiq@miun.se}}
\maketitle              
\begin{abstract}
In recent years, large language models (LLMs) have been extensively utilized for behavioral modeling, for example, to automatically generate sequence diagrams. However, no overview of this work has been published yet. Such an overview will help identify future research directions and inform practitioners and educators about the effectiveness of LLMs in assisting behavioral modeling.
This study aims to provide an overview of the existing research on the use of LLMs for behavioral modeling, particularly focusing on use case and sequence diagrams. 
Through a term-based search, we filtered and identified 14 relevant primary studies. Our analysis of the selected primary studies reveals that LLMs have demonstrated promising results in automatically generating use case and sequence diagrams. In addition, we found that most of the current literature lacks expert-based evaluations and has mainly used GPT-based models. Therefore, future work should evaluate a broader range of LLMs for behavioral modeling and involve domain experts to evaluate the output of LLMs.

\keywords{Conceptual modeling  \and Behavioral models \and Use case diagrams \and Sequence diagrams \and LLMS \and Large language models.}
\end{abstract}
\section{Introduction}
Unified Modeling Language (UML) has been widely used in software engineering to assist researchers, educators, and practitioners in requirement analysis and software modeling \cite{pilone2005uml}. For instance, UML has been used to create structural and behavioral diagrams that visualize the system's static architecture and describe its dynamic behavior, respectively. These diagrams support the understanding, communication, and documentation throughout the software development process. Generally, these graphical diagrams have been created manually, which is a time-consuming and human-intensive task. The advent of large language models (LLMs), such as ChatGPT\footnote{https://openai.com/index/chatgpt/}, Llama\footnote{https://www.llama.com/}, and Gemini\footnote{https://gemini.google.com/}, offers a promising solution for automated assistance in generating these graphical diagrams.

In recent years, LLMs have been extensively applied to various software engineering tasks, for example, in software testing \cite{hou2024large,wang2024assessing}, code generation \cite{gu2025effectiveness,jiang2024self}, and defect management \cite{laiq2023intelligent,laiq2024industrial,laiq2025comparative,laiq2025automatic}. 
Similarly, LLMs have been used for software design to assist in related tasks, for example, automating the creation of UML diagrams from natural language text \cite{garaccione2025evaluating,jahan2024automated}.

Several studies have applied LLMs to automatically create behavioral models (e.g., \cite{wang2024assessing,ferrari2024model,speth2024chatgpt}), including use case and sequence diagrams. However, no overview of this work has been published yet. Providing such an overview would help identify future research directions and inform practitioners and educators about the effectiveness of LLMs in supporting behavioral modeling. 

This study aims to fill the above-mentioned gap by offering an overview of the existing work on leveraging LLMs for creating behavioral models, with a particular focus on use case and sequence diagrams.

The remainder of this paper is organized as follows. Section \ref{sec:method} presents the study research method, including the search strategy, the study selection process, data extraction, and the analysis procedure. Section \ref{sec:results} presents the results of the study. Section \ref{sec:discussion} discusses the findings of the study. Section \ref{sec:conclusion} concludes the paper with future work.

\section{Research method} \label{sec:method}
To provide an overview of the existing work on the use of LLMs for behavioral modeling (i.e., use case and sequence diagrams), we conducted a literature survey. We followed similar guidelines as those followed by Taromirad and Runeson \cite{taromirad2025assertions}. This type of review is also labeled as a semi-systematic review \cite{snyder2019literature} and a systematic mapping study \cite{kitchenham2011using}. The process involves conducting a term-based search on Google Scholar, followed by a filtering process, and concluding with data extraction and analysis of the identified research.

\paragraph{\textbf{Search:}} To have a broad coverage of the relevant literature on the topic, we used Google Scholar as the primary search engine. We use the following two queries: ("large language models") AND ("UML sequence diagram") and ("large language models") AND ("UML use case diagram"). The search (performed on 29 June 2025) rendered about 59 and 54 hits for these two queries, respectively.

\paragraph{\textbf{Screening:}} The titles and abstracts were screened to select the relevant papers. As a result of the screening and removing duplicates, we identified 14 papers (see Table \ref{tab:selectedstudies}) on the use of LLMs for behavioral models (i.e., use case and sequence diagrams).

\paragraph{\textbf{Data extraction and analysis:}} Table \ref{tab:dex-template} shows the data extraction template that we used to extract data from the selected primary studies. We then categorized the studies based on the UML modeling tasks (i.e., use case and sequence diagrams), evaluated LLMs, used data sources, and evaluation approaches. We also provide descriptive statistics of the studied UML modeling tasks, evaluated LLMs, data sources, and evaluation approaches.

\begin{table}[!ht]
    \centering
    \caption{Data extraction template}
    \label{tab:dex-template}
    \begin{tabular}{p{1cm}p{8cm}}
    \toprule
     \textbf{Item} & \textbf{Description}
    \\ \midrule 
    
    \#1 & UML modeling task, e.g., generating a sequence diagram.
    \\ \midrule
    
    \#2 & Used LLM, e.g., GPT-4o.
    \\ \midrule
    
    \#3 & Used data source, e.g., use stories.
    \\ \midrule
    
    \#4 & Used evaluation approach, e.g., expert-based evaluation.
    \\  \bottomrule

    \end{tabular}
\end{table}

\begin{table}[!ht]
    \centering
    \small
    \caption{List of selected primary studies (PS)}
    \label{tab:selectedstudies}
    \begin{tabular}{p{0.8cm}p{11cm}}
    \toprule
     \textbf{Id} & \textbf{Title}
    \\ \midrule
    PS1 & Model generation with LLMs: From requirements to UML sequence diagrams \cite{ferrari2024model}
    \\ \midrule

    PS2 & Assessing UML Models by ChatGPT: Implications for Education \cite{wang2024assessing}
    \\ \midrule
    
    PS3 & Automated Derivation of UML Sequence Diagrams from User Stories: Unleashing the Power of Generative AI vs. a Rule-Based Approach \cite{jahan2024automated}
    \\ \midrule
    
    PS4 & How LLMs aid in UML modeling: an exploratory study with novice analysts \cite{wang2024llms}
    \\ \midrule
    
    PS5 & ChatGPT's Aptitude in Utilizing UML Diagrams for Software Engineering Exercise Generation \cite{speth2024chatgpt}
    \\ \midrule
    
    PS6 & Students-Centric Evaluation Survey for Exploring the Impact of LLMs on UML Modeling \cite{al2025students}
    \\ \midrule

    PS7 & Prompting Large Language Models to Tackle the Full Software Development Lifecycle: A Case Study \cite{li2024prompting}
    \\ \midrule
    
    PS8 & LLM-based System Design Automation (LSDA) Using Generative AI: Exploring a Formalized Framework Based on Experiments \cite{kumar2025llm}
    \\ \midrule
    
    PS9 & Evaluating Large Language Models in Exercises of UML Use Case Diagrams Modeling \cite{garaccione2025evaluating}
    \\ \midrule
    
    PS10 & Large Language Models Capabilities for Software Requirements Automation \cite{vega2024large}
    \\ \midrule
        
    PS11 & Empowering Software Architects with Artificial Intelligence: Analyzing GitHub Copilot's Role in Modern Architecture Design \cite{ramachandran2025empowering}
    \\ \midrule
    
    PS12 & Ai-driven consistency of sysml diagrams \cite{sultan2023clinical}
    \\ \midrule
    
    PS13 & Generative AI for Smart Contracts in Real Estate Business \cite{zhao2024generative}
    \\ \midrule
    
    PS14 & Exploring the Potential of Conversational AI Support for Agent-Based Social Simulation Model Design \cite{siebers2024exploring}
    \\ \bottomrule
    
    \end{tabular}
\end{table}

\section{Results and analysis}\label{sec:results}

\begin{table}[!ht]
    \centering
    \scriptsize
    \caption{A map of relevant studies on the use of LLMs for behavioral modeling (use case diagram (UCD) and sequence diagram (SD))}
        \label{tab:results}
    \begin{tabular}{p{0.8cm}p{1.5cm}p{3.5cm}p{3cm}p{3cm}}
    \toprule
        \textbf{PS \newline Id} & \textbf{Modeling task} & \textbf{Evaluated model} & \textbf{Data source (input)} & \textbf{Evaluation approach} \\ \toprule
        
        PS1 & SD & GPT‑3.5 & Requirement documents & Expert-based evaluation \\ \midrule
        
        PS2 & UCD,  SD & GPT‑4o & System description,  UML diagrams & Researchers as evaluators, LLM as a judge \\ \midrule
        
        PS3 & SD & GPT‑3.5 & User stories & Expert-based evaluation \\ \midrule
        
        PS4 & UCD,  SD & GPT-3.5,  GPT 4, NewBing & System description & Researchers as evaluators \\ \midrule
        
        PS5 & SD & GPT-4, DALL·E & System description, UML diagrams & Researchers as evaluators \\ \midrule
        
        PS6 & UCD,  SD & GPT‑4‑turbo & System description & Students as evaluators \\ \midrule
        
        PS7 & SD & GPT-turbo (3.5, 4),  DeepSeek-Coder (1.3B, 6.7B, 33B),  CodeLlama (7B, 13B, 34B) & Requirement documents & LLM as a judge \\ \midrule
        
        PS8 & SD & GPT‑4o & System description & Researchers as evaluators \\ \midrule
        
        PS9 & UCD & GPT‑4 & Requirement documents & Researchers as evaluators \\ \midrule
        
        PS10 & UCD& GPT‑4 & System description & Students/Researchers as evaluators \\ \midrule
        
        PS11 & UCD,  SD & GitHub Copilot & Software requirements specification & Similarity score measurement generated and referenced diagrams \\ \midrule
        
        PS12 & UCD & GPT-4-turbo, GPT-4o & System description & Rule-based consistency checker \\ \midrule
        
        PS13 & UCD,  SD & GPT-4o, Microsoft Copilot & Requirement documents & Researchers as evaluators \\ \midrule
        
        PS14 & UCD,  SD & GPT‑3.5 & System description & Expert-based evaluation \\ \bottomrule
        
    \end{tabular}
\end{table}

\begin{table}[!ht]
\centering
\caption{Descriptive statistics of UML modeling tasks, evaluated LLMs, data sources, and evaluation approaches.}
\vspace{-2mm}
\label{tab:results-analysis}
\begin{tabular}{cccc}
\\ \toprule
\textbf{Item}  & \textbf{PSs count} & \textbf{Item} & \textbf{PSs count} \\ \toprule
\underline{\textbf{\textit{Modeling task}}} &  & \underline{\textbf{\textit{Data source (input)}}} & \\
Sequence diagram   & 11    & System description  & 8 \\
Use case diagram    & 9  & Requirement documents   & 4  \\
\underline{\textbf{\textit{Evaluated model}}} & & UML diagrams  & 2 \\
GPT‑3.5 & 4 & User stories  & 1 \\
GPT-4 & 4 & Requirements specification & 1 \\ 
GPT‑4o & 4 & \underline{\textbf{\textit{Evaluation approach}}} & \\
GPT‑4-turbo & 3 & Researchers as evaluators & 7 \\
GPT‑3.5-turbo  & 1 & Expert-based evaluation & 3 \\
NewBing & 1 & LLM as a judge  & 2  \\
DALL·E & 1 & Students as evaluators & 2  \\
GitHub Copilot & 1 & Similarity score & 1  \\
Microsoft Copilot & 1 & Rule-based checker & 1 \\
DeepSeek-Coder  & 1  & &  \\
CodeLlama  & 1 & &   \\ \bottomrule

\end{tabular}
\end{table}

This section presents the results of our review on the use of LLMs for behavioral modeling (i.e., use case and sequence diagrams). Table \ref{tab:selectedstudies} shows the list of selected 14 primary studies on the topic. Table \ref{tab:results} shows a complete map of the identified studies, including the modeling task, the LLM used, the data source, and the evaluation approach.

Table \ref{tab:results-analysis} provides descriptive statistics of the identified primary studies about investigated modeling tasks, evaluated LLMs, data sources, and evaluation approaches. As shown in Table \ref{tab:results-analysis}, sequence diagrams were the most frequently investigated modeling task, addressed in 11 studies. Use case diagrams were also a significant focus, studied in nine studies. Several studies (9) focused on both diagrams, indicating a consistent interest in comparing the capabilities of LLMs across different behavioral modeling tasks.

Regarding the evaluated LLMs, as illustrated in Table \ref{tab:results-analysis}, GPT-based models are predominant. GPT-3.5, GPT-4, and GPT-4o were each evaluated in four studies. GPT-4-turbo was also commonly studied, appearing in three studies. Beyond these, other models included GitHub Copilot (PS11), Microsoft Copilot (PS13), NewBing (PS4), and DALL·E (PS5). Notably, PS7 assessed a range of open-source LLMs: DeepSeek-Coder (in configurations ranging from 1.3 billion to 33 billion parameters) and CodeLlama (7B, 13B, and 34B), reflecting a growing interest in comparing proprietary and open-source solutions for UML modeling tasks.

The data sources used as input to the models were diverse. According to Table \ref{tab:results-analysis}, the description of the system was the most common input, used in eight studies. Four studies relied on requirement documents, while two studies used UML diagrams as input prompts. A single study used user stories and software requirements specifications. This variety of input formats demonstrates the adaptability of LLMs in processing both structured and unstructured textual sources to generate diagrammatic representations.

As shown in Table \ref{tab:results-analysis}, the evaluation approaches varied considerably between the studies. The most prevalent method was human assessment by researchers, which was used in seven studies. Expert-based evaluation appeared in three studies. Two studies engaged students as evaluators, while two others relied on the LLM itself to judge the quality of its output. In addition, PS11 applied an automated similarity score to compare the generated diagrams with reference models, and PS12 employed a rule-based consistency checker to assess the validity of the output. This distribution highlights a continued reliance on human evaluation while also pointing to an emerging interest in scalable, automated assessment methods.

In summary, the studies presented in Table \ref{sec:results} demonstrate that LLMs have been widely applied to behavioral modeling tasks, particularly the generation of sequence diagrams, using a variety of models, input formats, and evaluation strategies. The emphasis on GPT-based models and the predominance of researcher-led evaluation reflect the current state of practice, while the inclusion of open-source models and automated assessment approaches suggests important directions for future research.

\section{Discussion}\label{sec:discussion}
In this study, we conducted a literature survey on the use of LLMs for behavioral UML models (use case and sequence diagrams). A total of fourteen primary studies were identified and analyzed. The results demonstrate that sequence diagrams were studied more frequently than use case diagrams, with eleven studies focusing on sequence diagrams and nine on use case diagrams. In the following, we discuss the findings of the survey and potential validity threats to it.

\paragraph{\textbf{Need for comprehensive evaluations of LLMs for UML behavioral models: }} Our findings indicate that among the evaluated LLMs in the literature, GPT-based models, including GPT-3.5, GPT-4, GPT-4o, and GPT-4-turbo, are dominant. Only one study (PS7) assessed LLMs other than those based on GPT, explicitly focusing on open-source models: DeepSeek-Coder and CodeLlama. Furthermore, most studies (nine) used only one LLM for their evaluations. Therefore, there is a pressing need for a comprehensive evaluation that includes a diverse range of proprietary and open-source LLMs. Conducting a comparative analysis is essential to understand the performance of different models. It will also facilitate trade-off assessments for their adoption in practice, as each model has advantages and disadvantages, such as operational cost differences.  

\paragraph{\textbf{Need for robust evaluation approaches:}} Our findings indicate that human assessment by researchers is a dominant evaluation approach among the evaluation approaches. Only a few studies (3) have used expert-based evaluations. While researcher-led assessments provide valuable insights into model accuracy and completeness, the predominance of subjective human evaluation indicates that the field remains in an exploratory phase. Therefore, more mature research is needed, ideally involving standardized rubrics, domain experts, and automated consistency checks to complement human reviews. Without such rigor, it remains difficult to draw strong conclusions about the effectiveness of LLM-based modeling tools.

\paragraph{\textbf{Implications for educators and students:}} In software engineering, various attempts have been made to assist students and junior engineers in their tasks through automated tools, for example, \cite{laiq2020chatbot,pirzado2024navigating,wang2024assessing,laiq2023data}. This review reveals similar attempts in software design, particularly in behavioral modeling. Since software architecture and design, including behavioral modeling, is often viewed as a complex subject for students \cite{nasir2022threshold}, automated support from LLMs can help lower barriers and improve learning outcomes. For example, students can use LLM-powered tools to receive feedback and reference examples, which helps accelerate their learning of UML modeling. They can iteratively improve their work by getting rapid evaluations of their diagrams. For educators, this review suggests that there are new opportunities. They can integrate LLMs as instructional aids by generating example diagrams, providing formative feedback on student submissions, or automating parts of the grading process. These tools can help reduce instructor workloads and minimize potential biases caused by fatigue or inconsistent evaluations.

\vspace{2mm}
\noindent \textit{In summary,} while the current body of work demonstrates that LLMs hold significant promise for automating behavioral UML modeling, important gaps remain in model diversity, evaluation rigor, and integration into practice. Addressing these challenges will be critical to realizing the full potential of generative AI in both education and practice.

\section*{Validity threats}

\textbf{\textit{(a) Missing of relevant studies:}} One potential threat to this survey could be that our search string misses the relevant primary studies on the topic. We mitigated this by carefully developing our search string based on the aim of the survey, i.e., LLMs for use case and sequence diagrams. In addition, to have a broad coverage of the topic, we used Google Scholar as our primary search engine without restricting our search to papers from specific years. 

\noindent\textbf{\textit{(b) Bias in paper selection and data extraction:}} Another threat to this survey is bias during the selection of papers and data extraction. We mitigated this risk by ensuring that every excluded paper was reconsidered after a one-week gap. Similarly, we rechecked all the data extracted. A similar test-retest approach is recommended for single reviewers conducting systematic reviews \cite{kitchenham2010s,petersen2011measuring}.

\section{Conclusion and future work}\label{sec:conclusion}

In this study, we conducted a literature survey on the use of LLMs for behavioral modeling, particularly use case and sequence diagrams. We identified and analyzed a total of fourteen primary studies. Our results indicate that sequence diagrams were studied more frequently than use case diagrams; eleven studies focused on sequence diagrams, while nine concentrated on use case diagrams.
Our findings indicate that most studies rely on the GPT family of models and incorporate human evaluation by researchers. Although the review suggests that LLMs have significant potential to automate behavioral UML modeling, there are still important gaps regarding model diversity, evaluation rigor, and the integration of these models into practice. Addressing these challenges will be essential for realizing generative AI's full potential in education and practice.

In the future, our goal is to comprehensively evaluate a diverse range of proprietary and open-source LLMs for behavioral modeling, including evaluating the output of LLMs with domain experts.

\begin{credits}
\subsubsection{\discintname}
The authors declare that they have no known competing financial interests or personal relationships that could have appeared to influence the work reported in this paper.
\end{credits}

\bibliographystyle{splncs04}
\bibliography{paper}

\end{document}